\magnification=1095
\input eplain
\input epsf

 
\ifx\tenpoint\undefined\let\loadedfrommacro=Y
\ifx\loadedfrommacro Y\else
         \message{10point.TeX must be loaded from a macro package.}
         \message{Input terminated.}
          \fi
 
\font\tencsc=cmcsc10
 
\newfam\scfam
 
\def\tenpoint{\def\rm{\fam0\tenrm}
    \textfont0=\tenrm  \scriptfont0=\sevenrm  \scriptscriptfont0=\fiverm
    \textfont1=\teni   \scriptfont1=\seveni   \scriptscriptfont1=\fivei
    \textfont2=\tensy  \scriptfont2=\sevensy  \scriptscriptfont2=\fivesy
    \textfont3=\tenex  \scriptfont3=\tenex    \scriptscriptfont3=\tenex
    \textfont\itfam=\tenit   \def\it{\fam\itfam\tenit}%
    \textfont\slfam=\tensl   \def\sl{\fam\slfam\tensl}%
    \textfont\ttfam=\tentt   \def\tt{\fam\ttfam\tentt}%
    \textfont\bffam=\tenbf   \scriptfont\bffam=\sevenbf
    \scriptscriptfont\bffam=\fivebf  \def\bf{\fam\bffam\tenbf}%
    \textfont\scfam=\tencsc  \def\sc{\fam\scfam\tencsc}%
    \normalbaselineskip=12pt
    \setbox\strutbox=\hbox{\vrule height8.5pt depth 3.5pt width0pt}%
    \normalbaselines\rm}

         \let\loadedfrommacro=N\fi
 
\font\ninerm=cmr9            \font\sixrm=cmr6
\font\ninei=cmmi9            \font\sixi=cmmi6
\font\ninesy=cmsy9           \font\sixsy=cmsy6
\font\ninebf=cmbx9           \font\sixbf=cmbx6
\font\ninesl=cmsl9           \font\ninett=cmtt9      \font\nineit=cmti9
\font\ninecsc=cmcsc10
 
\ifx\ninepoint\undefined
   \def\ninepoint{\def\rm{\fam0\ninerm} 
       \textfont0=\ninerm  \scriptfont0=\sixrm  \scriptscriptfont0=\fiverm
       \textfont1=\ninei   \scriptfont1=\sixi   \scriptscriptfont1=\fivei
       \textfont2=\ninesy  \scriptfont2=\sixsy  \scriptscriptfont2=\fivesy
       \textfont3=\tenex   \scriptfont3=\tenex  \scriptscriptfont3=\tenex
       \textfont\itfam=\nineit   \def\it{\fam\itfam\nineit}%
       \textfont\slfam=\ninesl   \def\sl{\fam\slfam\ninesl}%
       \textfont\ttfam=\ninett   \def\tt{\fam\ttfam\ninett}%
       \textfont\bffam=\ninebf   \scriptfont\bffam=\sixbf
        \scriptscriptfont\bffam=\fivebf   \def\bf{\fam\bffam\ninebf}%
       \textfont\scfam=\ninecsc  \def\sc{\fam\scfam\ninecsc}%
       \normalbaselineskip=11pt
       \setbox\strutbox=\hbox{\vrule height8pt depth3pt width0pt}%
       \normalbaselines\rm}
   \fi

\font\medbf=cmbx12
\font\sc=cmcsc10
\font\fignumfont=cmbxsl10 at 9pt

\advance\baselineskip by 1truept

\newcount\qnumber \qnumber=1
\def\nqlabel#1{\definexref{#1}{\the\qnumber}{problem}\nq}
\def\nq#1 \par{\medbreak\noindent\llap{\bf \the\qnumber~}{\it #1}%
\global\advance \qnumber by 1%
\par\noindent}

\newcount\footnotenumber \footnotenumber=0
\def\ft{\global\advance\footnotenumber by 1%
\footnote{$^{\the\footnotenumber}$}}

\def\part#1){\smallskip\noindent #1)}
\def\1{{\tt 1}}
\def\ans{{\it Answer:\/}\ }
\def\u#1{\,{\rm #1}}
\def\s{\,{\rm s}}
\def\m{\,{\rm m}}
\def\myfig#1{\medskip$$\vbox{\halign{##\hfil\cr\vbox{\epsfbox{#1}}\cr}}$$}
\def\t{\theta}
\def\xeqno#1{\eqdef{#1}}

\def\w{\omega}
\def\half{{1\over2}}
\def\x{\times}

\def\e#1{10^{#1}}

\listrightindent=15pt
\interitemskipamount=1.5pt

\newcount\sectionNum \sectionNum=0	
\newcount\subsectionNum \subsectionNum=0

\def\newsectionNum{\global\advance\sectionNum by 1 \global\subsectionNum=0
\relax \thesectionNum\relax}
\def\newsubsectionNum{\global\advance\subsectionNum by 1 \relax
\thesubsectionNum\relax}

\def\thesectionNum{\the\sectionNum}
\def\thesubsectionNum{\the\sectionNum.\the\subsectionNum}

\def\section#1 \par{\bigbreak\noindent
{\bf \llap{\newsectionNum~}#1}\nobreak\smallskip\noindent\ignorespaces}

\def\sectionl#1#2 \par{\section{#1} \par
\definexref{#2}{\thesectionNum}{section}}

\def\subsection#1 \par{%
\bigbreak\noindent
{\it \llap{\newsubsectionNum~}#1}\nobreak\vskip1.5pt\noindent\ignorespaces}

\def\subsectionl#1#2 \par{\subsection{#1} \par
\definexref{#2}%
{\thesubsectionNum}{section}}

\newcount\figureNum \figureNum=0
\newbox\figbox \newdimen\xsize \newskip\ysize \newbox\realfigbox
\newbox\figboxtemp \newbox\tempbox

\def\ifig#1#2#3{%
\global\advance\figureNum by 1%
\definexref{#2}{\the\figureNum}{figure}%
\midinsert
\parindent=0pt
\halign{##\hfil\cr
\leftline{\hskip\leftskip\vbox{\epsfbox{#1}}}\cr \noalign{\smallskip}
\vbox{\hsize=5truein%
\noindent {\fignumfont \noindent Figure~\the\figureNum.}~#3}\cr}
\par
\endinsert}

\raggedbottom
\newdimen\size
\def\lengthen#1{\size=#1\advance\vsize by \size \advance\voffset
by -0.5\size} 

{\parindent=0pt \rightskip=0pt plus 1fill
{\medbf Observations on teaching first-year physics}
\footnote{\null}{Copyright \copyright\ 1998--2005 by
Sanjoy Mahajan.  Licensed under the Open Software License version 3.0.
See the file COPYING in the source code.}
\medskip
\obeylines
{\sc Sanjoy Mahajan}
Cavendish Laboratory
Astrophysics
University of Cambridge
Cambridge CB3 0HE
England
{\tt sanjoy@mrao.cam.ac.uk}
}

\bigskip\hrule\smallskip
\noindent {\bf Abstract.} 
Highly successful students, as measured by grades and by scores on the
Force Concept Inventory, still struggle with fundamental concepts in
mathematics and physics.  These difficulties, which turn physics into
parrot learning and include confusing velocity and acceleration or
being unable to reason with graphs, are revealed by problems requiring
estimation and conceptual reasoning.  I discuss these problems, the
difficulties that they reveal, and suggest possible remedies.
\smallskip\hrule\bigskip

\noindent
I gave tutorials to ten students taking the first-year (IA) physics
course at Cambridge University.  The students -- diligent, curious,
and a joy to teach -- had studied physics in high school for years
and, as measured by grades and by scores on the Force Concept
Inventory, with great success.  However, using problems requiring
estimation and conceptual reasoning (collected in the Appendix), I
found that they struggle with fundamental concepts in mathematics and
physics.  These difficulties -- such as confusing velocity and
acceleration or being unable to reason with graphs -- prevent them
from understanding or appreciating the beauty of physics, and force
them into rote or parrot learning.  Physics becomes a game of memory
and formula juggling.  We can avoid this disastrous result by teaching
students how physicists think: how we approximate and reason in
unfamiliar situations.
\ref{sec:suggestions} contains suggestions in this direction.  

\section{Physics difficulties}

Students live in the pre-Newtonian world and do not understand
acceleration; they confuse Newton's second law and third laws; they
find circular motion confusing; and they cannot make or reason with
freebody diagrams (diagrams of one object -- the free body -- in which
every other object is replaced by a force on the free body).

Such difficulties go unnoticed because students can solve many
standard problems in spite of the difficulties; they are talented and
have memorized rules that are often true.  For example, students know
that in circular motion some force will be $F=mv^2/r$, if only because
that formula is highlighted in the textbook section on circular
motion.  They are not sure of the force's direction or cause,
but problems often specify $F$, $m$, and $r$, and ask for $v$.  Simple
algebra yields $v$, whether or not the student understands the cause
or direction of the force.


\subsection{Aristotelian thinking: The Force Concept Inventory}

Perhaps the most fundamental physics misconception is confusion
acceleration with velocity: Students believe that zero velocity
implies zero acceleration and therefore zero net force.  This belief
is an example of Aristotelian, or more precisely, of pre-Newtonian
thinking.

Such beliefs are tested for by the Hestenes--Halloun Force Concept
Inventory (FCI) \cite{Hestenes-1992}, which contains 30
multiple-choice questions requiring no calculation but rather a solid
understanding of Newtonian principles.  For example, one question asks
about a truck (lorry) colliding with a small car: How does the force
on the truck from the car compare with the force on the car from the
truck?  The typical answer, which reveals that students do not
understand the third law, is that the truck exerts more force than the
car exerts.  Many American physics departments use the FCI before and
after a semester of first-year mechanics (which is roughly at the
level of British A-level physics).  Typical pre-instruction scores for
American students range around 50\% but that number varies greatly
among universities.  Hestenes and Halloun state that scores of 60\%{}
indicate entry into Newtonian thinking, and that scores above 85\%{}
indicate `confirmed Newtonian think[ing]' \cite{Hestenes-1995}.  I
gave the FCI to my ten students as their first assignment before
lectures began.  The lowest score was 25/30 (83\%) and the average was
27.5/30 (92\%); one student had a perfect score of 30/30.  The
students did extremely well on this measure.

\subsection{Beyond the Force Concept Inventory}

Yet I doubted this conclusion for a number of reasons.  First, on the
difficult questions (such as the truck question), students who
answered it correctly often circled and then scribbled out one or two
wrong answers before choosing the correct answer.  Second, I realized
how late in my study of physics I had sorted out the difference
between Newton's second and third laws: only when I helped to teach
first-year physics as a graduate student.

The FCI is too easy for Cambridge students; they study physics and
mathematics for many years in high-school, far more than American
students: In high school the Cambridge students take A-level
Mathematics, Physics, and Further Mathematics (and often Chemistry).
Each A-level course, which takes up one-third or one-fourth of a
student's time for the last two years of high school, is equivalent to
perhaps 1.5 Advanced Placement exams.

So, as part of the weekly tutorial work, I assigned free-response
problems that required Newtonian thinking.
\ref{prob:skiing-pendulum}, for example, asks about skiing downhill holding a
pendulum (the problems are collected in the Appendix).  A few students
correctly guessed that, without friction, the bob in case~1 points
perpendicular to the hill.  I say `guessed', because none produced a
freebody diagram or other convincing argument.  All except one student
said that the pendulum points directly downwards in case~3, when the
skier is motionless at the top of her path.  One student realized that
in each case the pendulum points perpendicular to the hill; he was the
only student to realize that velocity was independent of acceleration.
He also had the only perfect score on the FCI.

\ref{prob:steel-ball}, about a bouncing steel ball, brought out the same
difficulty.  The freebody diagrams were correct for cases~1 and~2, but
were wildly incorrect for case~3 (when the ball is motionless as it
rebounds).  A typical diagram had a downwards gravitational force of
$mg$ balanced by an upwards `reaction' force of $R=mg$.  The most
common justification for $R=mg$ was that the velocity is zero, so the
body must be in equilibrium; another justification was that most
reaction forces met in A-level problems were equal to $mg$ (see
\ref{sec:parrot-learning}).  This `zero velocity implies zero force'
reasoning illustrates the pre-Newtonian belief $F\propto v$, rather
than the Newtonian law $F=ma$.  One student realized that the upwards
force had to be larger than $mg$, otherwise the ball would never leave
the ground, but even he said that $R$ was only $2mg$.  In the tutorial
we estimated the upwards force by modelling the steel ball as a
springy cube; all the students were surprised to find that
$R\sim10000mg$.  To get a feel for these magnitudes, students could
study such systems in their laboratory courses.

\subsectionl{Newton's second and third laws}{sec:N2-N3}

Students also have difficulty distinguishing Newton's second from
Newton's third law.  Most of the difficulty results because students
do not understand the third law.  \ref{prob:scales} asks students to
prove that a composite object has weight equal to the sum of the
individual objects' weights (for a two-object system).  None of the
students provided a proof, and their explanations confused the second
and third laws.  I therefore assigned the problem again, giving more
instructions, as \ref{prob:scales-again}.  (The difficulty with proof
is a mathematical trouble, and the topic of \ref{sec:proof}.)

Students stated that the force of their hand on the book equals the
weight of the book, `because of action--reaction'.  They did not
realize that they had implicitly invoked equilibrium and must
therefore use Newton's second law to conclude that the two forces {\it
on the same object\/} are equal and opposite.  Students were surprised
to find that the gravitational force of the book on the earth is the
third-law counterpart to the weight of the book.  They do not
understand the third law as a statement about interactions, so they
see any pair of equal and opposite forces as a third-law pair.  I
often asked students to discuss a law for a candidate force:
$$F=k {s_1 s_2^3\over r^4},$$ where $s_1$ and $s_2$ are charges,
analogous to mass or electric charge, and $k$ is the constant that
makes the dimensions correct.  Could such a force exist?  Students are
pleasantly surprised that the third law forbids such forces because
its force pairs are not equal and opposite.

\subsection{Heavier objects fall faster}

The classic Aristotelian belief is that heavier objects fall faster
than lighter objects.  Surely three hundred years after Galileo showed
otherwise, students no longer share this belief?  Unfortunately, many
do, but the belief shows up only in novel situations.  Students know
that if a stone and a cannonball fall, they should {\it say\/} that
both objects hit the ground `at the same time'; if they have been
carefully taught, they might even say `roughly at the same time'.
They also know what to say about two objects sliding down an incline,
that mass is irrelevant.  However, when the problem includes the novel
effect of rolling (yet more trouble with circular motion!), many
students have no practiced Newtonian answer to quote, and reveal
their gut-level Aristotelian belief.  For example, in
\ref{prob:rolling}, about objects rolling down a plane, some students
reasoned that an object with a large moment of inertia, such as a
disc, rolls faster than an object with a small moment of inertia, such
as a solid sphere.  Two students argued that `moment of inertia is
analogous to mass, and heavier objects fall faster than lighter
objects'!  I could not agree with the analogy, but I admired its
boldness.

The way that rehearsal hides this misconception reminds me of the
theory of the English accent: that if you step on an Englishman's toes
in the middle of the night, he'll shout at you in an American accent.
On this view, the one true accent is American.  An English accent is
just an act, a mask dropped upon surprise.  Similarly, the students'
response that `all objects fall at the same speed' is carefully
rehearsed.  It falls away when we step on their toes by asking about
it in a novel context, whereupon they reveal their true belief, that
heavier objects fall faster.

\subsection{Centrifugal force}

Students think that circular motion implies a centrifugal force.  I
asked students to draw a freebody diagram for an amusement park ride.
In this ride, you stand against the edge of a cylindrical cage that
spins rapidly; eventually the floor drops away.  But, voil\`a, you
remain against the wall.  This ride is the subject of a typical
high-school physics problem: Find the angular velocity such that a
person does not slide down the wall.  Students solve it correctly,
because they need only know that some radial force is $F=mv^2/r$
(never mind in what direction it points) and that the friction force
is $\mu F$.  However, when they draw a freebody diagram, their
confusion is evident.  A typical diagram is \ref{fig:centrifugal}.
Students insert the centrifugal force, because some force is `throwing
the person outwards (which is why you feel pressed against the wall).'

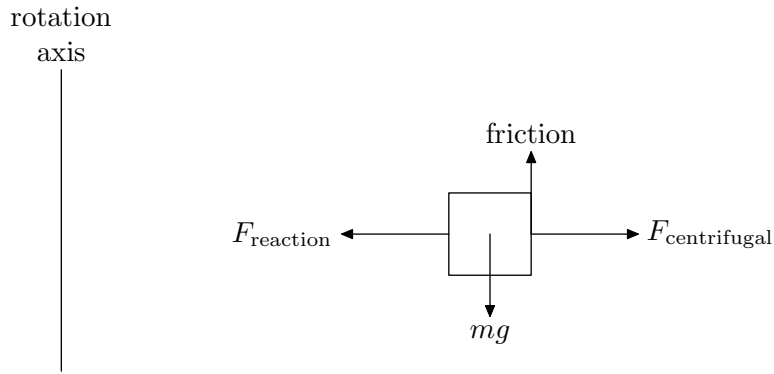
\ifig{fig.5}{fig:centrifugal}{Freebody diagram for amusement park ride.}

\subsection{Freebody diagrams}

The trouble with acceleration and confusion about third-law pairs
means that students cannot make freebody diagrams.  In answers to
\ref{prob:scales} (standing on a scale with a book in hand), most students
drew book, person, and scale with no separations, and drew ambiguous
contact forces on the border between objects.  These difficulties are
typical when students first learn freebody diagrams
\cite{Arons:1997}.  I had to explain that freebody diagrams
are diagrams of one object (or of one composite object) -- the free
body -- with other objects replaced by a force on the free body.

If students understood this replacement principle -- and the idea of
{\it system} -- they would not double count by inserting centrifugal
forces.  Only when I asked them what object causes the centrifugal
force did they realize that this force merely labels an actual force
and has no separate existence.  Unlike experienced physicists,
students do not naturally make freebody diagrams to analyze confusing
situations.  None of the students made a freebody diagram for
\ref{prob:skiing-pendulum} (skier holding a pendulum), even the
students who drew the correct pendulum positions.  In university
physics we need to teach this valuable skill, which is not part of
most British school physics curricula (although it is in most American
physics textbooks).

\section{Mathematical troubles}

Students have many mathematical difficulties.  They have not been
taught how to construct proofs or how to make educated guesses.  They
cannot make numerical estimates or reason using graphs.

\subsectionl{Proof and guessing}{sec:proof}

In \ref{sec:N2-N3}, I alluded to students' difficulty with proof
(\refs{prob:scales} and~\refn{prob:scales-again} on weighing a
composite object).  The solutions had numerous holes, besides the
errors in using Newton's laws.  When confusion between Newton's second
and third laws barred legitimate progress, students assumed the
conclusion.  Perhaps school mathematics should reintroduce Euclidean
geometry, not to indoctrinate students with 10,000 theorems about
triangles, circles, and diagonals of rhombuses, but to teach proof.
We want students to learn how to distinguish sound from unsound
arguments, whether or not they become mathematicians.

A complementary difficulty is fear of guessing.  Students have not
been taught techniques for making educated guesses
\cite{Polya:1954-1}; they are therefore reluctant to guess a solution before
solving a problem exactly.  Having have no feel for how a result
should turn out, they instead push symbols around until a reasonable
formula appears and declare it to be the answer.

\subsection{Heuristic arguments}

Students do not know how to tell whether an answer is reasonable.  For
example, they do not naturally use heuristics such as checking
limiting cases, or use more rigorous methods such as dimensional
analysis.  This difficulty is related to their reluctance to guess: If
students had a clear idea of what answers might be reasonable, they
would find it easier to guess an answer.

As practice with heuristic methods, I assigned \ref{prob:range}, analyzing
the formula for the projectile range.  A few students realized that,
for example, $v^2$ results from one $v$ in the flight time and one $v$
in the horizontal speed.  Many, however, refused to make a heuristic
analysis, and instead derived the range formula in the standard way.

Almost everyone is reluctant to make guesses, especially in a
supposedly exact subject such as physics.  To overcome this natural
reluctance, we must teach students heuristic methods; with practice,
students will develop the courage to use them.

When I explicitly forbade messy calculations and forced students to
use intuitive arguments, many were stumped.  \ref{prob:moments} asked
them to determine, without evaluating any integrals, whether a
spherical shell or a flat disc (same radius and mass) has the larger
moment of inertia.  Only one student found a correct argument:
squashing the sphere perpendicular to the axis of rotation, and
comparing the squashed mass distribution to the mass distribution of
the flat disc.  Some evaluated the integrals, in spite of the
instructions.  But many evaluated the wrong integral, $\int \rho
r^2\,dV$, rather than $\int
\rho(x^2+y^2)\,dV$ (for rotation about the $z$-axis).  More practice with
heuristic arguments, leading to conceptual understanding, would help
them set up the correct integral.

\subsection{Graphical reasoning}

Students cannot reason using graphs.  The troubles are mathematical
and physical.  A mathematical trouble is shown in the answers to
\ref{prob:Stirling}a (deriving Stirling's formula).  Students 
correctly drew a diagram like \ref{fig:Stirling}.
They then had to decide, `Does the integral over- or
underestimate the sum?'
Even with the clear graph {\it that they had drawn},
they did not see that the area
under the smooth curve
is less than the area under the total rectangles.  Instead,
this realization
came only by using a calculator to evaluate $$\sum_1^7\ln k=8.52\ldots$$
and finding it
larger than $$\int_1^7\ln k\,dk=7\x(\ln7-1)+1=7.62\ldots$$
I cannot quarrel with the answer but the method reveals serious
difficulty reasoning with graphs.  In a famous example,
Wertheimer \cite{Wertheimer:1959} would ask school students 
questions like
$${273+273+273+273+273\over5}=\,?$$
Some students got the joke and laughed.
Some protested that the calculations were too hard.  Most worryingly, some
added the 273's and then used long division to
find\dots273.  With the answer one has no quarrel, but like with the
logarithm graph, the method reveals a fundamental difficulty.

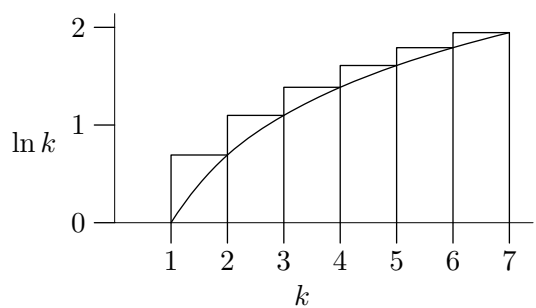
\ifig{fig.1}{fig:Stirling}{Deriving Stirling's formula.}

Students also do not realize the physical implications of a graph.
For example, in \ref{prob:snooker} -- about a snooker ball slipping and
rolling -- they realized that the ball slips for a while, then
eventually rolls, although not all correctly solved for the time $t_0$
until pure rolling.  However, their graphs of center-of-mass velocity
$v$ and of scaled angular velocity $r\w$ looked like
\ref{fig:rolling}.  Students did not realize that the rolling
condition, $v=r\w$, requires that the two curves coincide after $t_0$.  This
difficulty is also related to confusion about rolling motion.
Students have memorized $v=r\w$, but are not sure of its character.
They wonder if it is the definition of $\w$.  It took much discussion
to convince them that $v=r\w$ follows from the definition of rolling:
Motion where the point of contact is motionless.  This difficulty with
the character of $v=r\w$ is related to the difficulties with proof.
Students rarely know what assumptions a formula requires, or what
consequences follow from a formula.

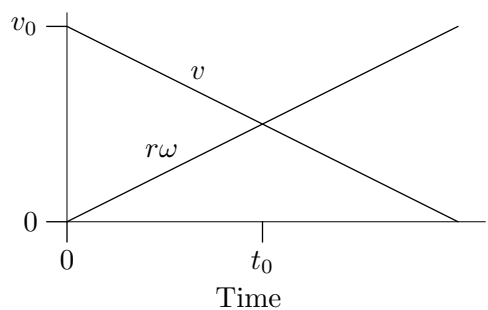
\ifig{fig.2}{fig:rolling}{Snooker ball rolling and sliding.}

\subsection{Number sense and approximations}

Perhaps the most serious mathematical difficulty is an atrophied
number sense and an inability to approximate.  For example, I asked
students to estimate $\sqrt{1.2}$, and most had no idea.  After I
prodded them to guess anyway, one said $1.05$; one said $1.\sqrt{2}$
meaning $1+\sqrt{2}/10$ (at least a creative answer).  Only one
student said $1.1$, and he was unsure of himself.  All the students
except the one who answered correctly had been using calculators
since age 11.  This other student, who went to school in
Singapore, had not been permitted to use a calculator until the last
two years of high school.  Ironically, some students who could not
evaluate $\sqrt{1.2}$ could instantly tell me that $\sqrt{1+x}$ is
roughly $1+x/2$ by the binomial theorem.

I am not surprised by this lack of number sense.  I co-taught a short,
intensive physics course to twenty A-level students; the course is run
by Villiers Park, a charity in Foxton.  The students, from all over
the United Kingdom, were in their final year of high school, taking
A-levels in Physics, Mathematics, and Further Mathematics.  Each
participating school nominated its best physics student, and these
students were all talented and curious about physics.  Over one-half
had interviews at Cambridge for admission to the Natural Sciences
degree (which includes the physics major), and a significant fraction
of them will attend Cambridge to study physics.  Yet none could
estimate $\sqrt{1.2}$; one student had trouble working out $120/10$.

Following up on approximations, I asked my Cambridge students to
estimate $\ln1.25$.  Some said that there was a series for $\ln$,
which they could not remember.  Some remembered that $\ln(1+x)\approx
x$, but could not say why the approximation was plausible.  A
graphical approximation using the definition
$$\ln (1+x) \equiv \int_1^{1+x} {dx\over x},$$ was a new and pleasant
surprise for them.

\subsection{Mathematics obscuring physics}

Mathematics in a derivation often prevents students from understanding
physics in the derivation.  For example, in the kinetic-theory
derivation of the diffusion coefficient, a series of flux integrals
simplify to
$$D={1\over3}\ell c,$$ where $\ell$ is the mean free path and $c$ is
the root-mean-square particle velocity.  Students do not realize that
all the fiddling with integrals of sines and cosines gives only the
factor of $1/3$, and that the factor of $\ell c$ is independent of the
angular integrals.  Partly, they have not been taught dimensional
analysis, which requires that $$D=(\hbox{dimensionless
constant})\x\ell c.$$ Partly, they do not understand diffusion even in
one dimension, where angular factors no longer torment them.  We
should spend most of the time on the qualitative reasoning in one
dimension, and assert the result for three dimensions by fiat, giving
the derivation in Eric Rogers' classic text, {\it Physics for the
Inquiring Mind\/} (Princeton University Press, 1960).  That
derivation, also used in the old Nuffield O-level physics course (for
ages 11-15!), uses six swarms of molecules, each marching along one
coordinate direction.

\sectionl{Parrot learning}{sec:parrot-learning}

One theme has shown up in many of these examples: A-level learning is
parrot learning.  Students memorize rules without understanding their
origin or range of validity.  As Mark Twain said, `It's not what I
do not know that worries me, it's what I know that ain't so.'
Here are a few `not-so' stories, culled from what students
believe:
\unorderedlist
\def\i#1. {\li {\it #1.}\ }

\i Every normal or reaction force is $mg$ (or, on an incline,
$mg\cos\theta$).  This rule is often true, but students do
not know when, and invoke it as a law of physics.
\ref{prob:skiing} asks what a scale reads when the scale--person system
accelerates down a plane.  A few students correctly reasoned that it
reads $3mg/4$, but most concluded that the scale reads $mg$.

\i In circular motion, there is a centrifugal force $mv^2/r$.  Such a
force exists in the reference frame of the
moving object, but students use it even in the laboratory frame, and
do not realize that its validity depends on the frame.

\i Tension is a force.  This rule is always wrong; students induce it
after seeing many freebody diagrams in which arrows representing
forces are labelled $\bf T$.  I once believed the rule for the same
reason.  Now I always label forces produced by tension as ${\bf
F}_{\rm T}$.  The rule causes trouble when, for example, students
analyze a tug-of-war. Two people hold one end each of a rope and pull
with force $100\u{N}$; what is the tension in the rope?  Some students
say zero: They add the two end forces (to get an alleged force, the
tension).  Some say $200\u{N}$: They add the force vectors as if they
were unsigned scalars.  A few say $100\u{N}$, without conviction.  We
need not explain tension in its full glory as a component of the
stress tensor, but we need to eradicate this rule.

\i In oscillating motion, $\hbox{acceleration}=0$ at the
equilibrium position.  This rule is correct for simple harmonic
motion, but students apply it too widely.  For example, in
\ref{prob:pendulum}, asking about pendulum motion, most students stated that
the bob was not accelerating at point~C, probably because they had
memorized pendulum motion as an example of simple harmonic motion, and
did not pause long enough to think that the bob must have an inwards
acceleration to move in a circle.  This difficulty is not confined to
first-year physics students.  At the University of Washington, 0 out
of 120 first-year physics students answered it correctly, a result
that no longer surprises me; however, only 15\%\ of students taking
their PhD qualifying exams answered it correctly \cite{Reif}!

\ifig{fig.3}{fig:perpetual}{Perpetual motion.}

\i Buoyancy can be replaced by an upthrust, a vertical force with
magnitude equal to the weight of fluid displaced.  This rule is often
true, but not in this perpetual-motion machine of
\ref{fig:perpetual}. The figure shows a cross-section of the machine:
The circle is a long cylinder, and the dot at its center is a spindle
that allows it only to rotate, not translate.  The thick vertical line
is a barrier that prevents the mercury and water from mixing.  I
explain to the students why the fluid exerts a torque: The upthrust
from the mercury side (longer arrow) is 13 times larger than the
upthrust from the water side (shorter arrow), because mercury is 13
times denser than water.  So the spindle rotates: perpetual motion!
None of the students exorcised the perpetual-motion demon without
extensive help, because they do not understand how upthrust or
buoyancy is a consequence of pressure forces (which in this case all
act through the spindle and hence provide no torque); instead,
upthrust or buoyancy is a memorized word.

Even when the upthrust rule is true, students do not realize its
origin.  For example, in \ref{prob:isothermal-atmosphere},
analyzing the isothermal
atmosphere, students counted the buoyant force twice in their freebody
diagram for a slab of air (\ref{fig:air-slab}).
In the figure, $A$ is the bottom or top surface area of the slab,
and $P_1A$ and $P_2A$ are the pressure forces.  Students realized that
$P_1>P_2$, and solved correctly for the pressure versus height, but
did not realize that the pressure forces already included the
upthrust.

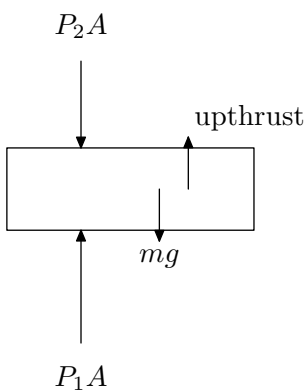
\ifig{fig.4}{fig:air-slab}{Incorrect freebody diagram for a slab of
air.  The upthrust duplicates the result of the two pressure forces.}

\noindent
Many school physics courses do not include Archimedes' principle; those
that do often simplify the treatment to `upthrust', with no discussion
of its origin in pressure forces.  The preceding examples show the
danger of such a simplification, which provides little scope for
thoughtful physical reasoning.

\endunorderedlist

\noindent Parrot learning makes physics into a memory game, and
students see physics the way many see history: as a collection of
facts to memorize.  Professional historians are repelled by this view
of history, as we are by the same view of physics.  Doing and enjoying
physics requires understanding fundamental principles and seeing how
they connect with one another.

\sectionl{Suggestions}{sec:suggestions}

After years of school physics, students should not have the
mathematical and physical misconceptions that I have discussed.  When
they come to university, we should be able to discuss problems and
ideas with them as budding physicists, even if they later specialize
in other subjects.  We obviously do not live in such a world;
high-school physics does not give students a high level of
mathematical and physical understanding.

We cannot expect any improvement soon.  On the contrary, most changes
in the school curriculum increase students' reliance on calculators
and reduce the physics and mathematics that they must know.
Furthermore, many teachers, products of the school and university
physics-teaching system (we share a lot of responsibility for the
problem), have some of the above misconceptions; every time I teach, I
find misconceptions in my own thinking.  How can students learn what
their teachers do not understand and what their textbooks do not
mention?

Instead of compounding the misconceptions, as we traditionally do in
university physics teaching, we could remedy the defects.  One
solution is to teach qualitative physics.  By qualitative I do not
mean physics for poets \cite{March:1995}; it is an excellent idea for
a course, but it might poorly serve future scientists.  Rather, I mean
that we teach dimensional analysis, heuristic methods, graphical
reasoning, and the arts of approximation and guessing: We should teach
students how physicists think.

We can illustrate these methods with applications to everyday physics;
for example, to stirring tea (\ref{prob:teacup-spindown}).  Students
are fascinated by such problems.  When I assigned the tea problem,
they gathered in each others' rooms and spent hours stirring tea and
timing the spindown.  Such problems give students a graspable example
of a physical concept (in this case, diffusion of momentum).  Using
everyday examples, students get feedback from the world on the
correctness of their physical picture. When students study waves and
oscillations, they can apply their knowledge to the physics of music,
a subject that interests most physics students.  Such an approach will
inspire students and encourage them to think like physicists.

I present here a few methods to teach qualitative physics, in order of
increasing headache to implement.  A few methods apply more to the
British university system, but I have tried to make most methods of
wide applicability.

\subsection{Peer instruction}

Eric Mazur at Harvard developed a simple method for getting students
to think qualitatively: {\it peer instruction\/}
\cite{Mazur,Mazur:1997}.  After explaining a
concept, such as buoyancy, he stops and puts on the overhead projector
a multiple-choice question -- called a Concept Question -- for
students to answer individually.  The question is easy for the student
who understands the principle; otherwise it takes a while, longer than
Mazur gives them.  One buoyancy question is:
\smallskip
{\advance\leftskip by \parindent \advance
\rightskip by \parindent \noindent 
Two cups are filled to the same level with water. One of the
two cups has ice cubes floating in it. Which cup weighs more?
\numberedlist \interitemskipamount=0pt \listcompact
\li The cup without ice cubes. 
\li The cup with ice cubes. 
\li The two weigh the same.
\endnumberedlist\par}
\noindent
To allow no time for useless calculation, Mazur gives students only
two minutes.  Then he asks students which choice they picked.  After
this public commitment, each student spends one or two minutes
convincing her neighbor of her answer -- the key to Mazur's method.
In explaining their choice, students realize what concepts confuse
them and begin to sort out their confusions.  And they get interested
in the material as they defend their views.  The discussion improves
their attention and their intuition.

Mazur breaks lecture into 15-minute blocks; each block has a short
explanation and then time for a Concept Question.  But even one
question per one-hour lecture (the format used in Caltech first-year
physics course) improves students' attention and understanding.  Peer
instruction has several merits.  First, it requires no fancy hardware
in the lecture theatre (although Mazur's classroom has networked
palmtop computers for the students to enter answers); I get students
to close their eyes (to prevent the herd effect) and vote by raising
hands.  Second, anyone can try it, using either their own questions or
the database of {\it ConcepTests\/} in Mazur's book.

\subsection{Two-week intensive preparation in the summer}

Another possibility is to offer a two-week intensive
`order-of-magnitude physics' course for students before they start
their year of physics.  Two weeks of intensive teaching is enough time
to teach the main ideas, especially if the rest of the year
occasionally uses the ideas taught in the intensive course.  At
Villiers Park, I taught qualitative physics for half a week to
students in their last year of high school.  The students enjoyed it,
and by the end of the session, after they had seen the principles
illustrated with many examples, they grasped the main ideas.

\subsection{Vacation study}

Or, the regular teaching could remain mostly as it is and instead
students could learn qualitative physics during the breaks between
terms.  This approach applies especially to British universities with
their short terms and long breaks (especially to Oxford and Cambridge,
where terms are only eight weeks!).  With this approach, the
examination at the end of the first year should contain questions that
require such reasoning, otherwise students might spend the entire
vacation recovering from sleep deprivation rather than also learning
physics.  Students would need to written material to learn from,
ideally a textbook on approximation and based on the first-year
physics topics.  The vacation-study approach has pros and cons.  On
the bad side, it reduces their sleep.  Perhaps more fair is to winnow
the standard topics, and use the time saved to teach approximation
during the year.  On the good side, it encourages students to learn
from textbooks, a skill valuable especially after they finish their
degree.

\subsection{Alter tutorials}

In the Oxford and Cambridge system, with tutorial as well as lecture
teaching, the lectures could remain traditional while tutors could
teach qualitative physics.  In the American system, the sections could
teach qualitative physics, leaving lectures alone.  As with the
vacation-study approach, the exam would need to be changed to
emphasize the value of qualitative reasoning.  Many graduate students,
who are a large fraction of the tutors or section leaders, do not feel
confident teaching material that they did not learn at university.
They would need training.

With proper training, this approach can work well, even if it is used
only for one term.  I used it with my students, assigning them the
problems in the Appendix and using tutorials to discuss the
difficulties.  The students and I enjoyed these problems.  They
prepare students to think like physicists, although alone they do not
prepare students for the first-year exam.  So I asked students to use
the Christmas vacation to practice old exam problems on the first
term's material.  The students were sufficiently happy with the method
to do as I asked, but it requires extra time from them and their
tutors.

\subsection{Modify lectures and tutorials}

The first term, or the first year, could teach qualitative physics --
in lectures and tutorials.  On the down side, this approach combines
the problems of the alter-tutorials approach (training tutors) with
the pain of redoing the lectures.  A specially written textbook would
be useful here.  This approach, although painful, has the best chance
of teaching the physics and mathematics that we want students to know.

Even with a radical approach of devoting the entire first year to
qualitative physics, students would not be harmed by the deemphasis on
exact calculations.  Those continuing as physics majors will practice
exact analyses in their second and third years; by then their
mathematical maturity will be greater and the analyses will not hinder
their understanding of physics (what it does in the first year).
Students majoring in chemistry, geology, or material science, who will
study only the qualitative physics, will also benefit.  A geologist,
for example, needs to estimate the relative contributions of
convection and conduction in transporting heat in the mantle more than
she needs to solve exactly a model that includes only conduction.  In
general, non-majors need intuitive understanding of physics more than
they need exact calculations.

\medskip
\noindent The difficulties that students have with physics and mathematics
are soluble.  Using the methods above, I hope that we can introduce
students to many years of understanding and enjoying physics.

\bigskip\bigskip
\leftline{\medbf References}
\bibliography{report}
\bibliographystyle{plain}

\vfill\eject
\leftline{\medbf Appendix}\bigskip
\noindent I assigned these problems to my
students in the first term of their first year.  
\medskip


\nq Estimations

\part a) How many English words can you recognise?

\part b) How many pieces of mail does the UK postal system carry each
day?  Estimate the annual budget of the Royal Mail; check your
estimate by looking up a recent Royal Mail budget.

\nqlabel{prob:range}Interpreting equations

Here you will study the well-known formula for the horizontal range of
a rock.  You throw a rock with velocity $v$ at an angle $\t$ with
respect to the ground.  Its range is $$R={2v^2\over
g}\sin\t\cos\t.\xeqno{range}$$ You can increase your confidence in
this result in a number of ways (parts a--e).  (It may help for many of
the parts to draw a diagram.)

\part a) Dimensional analysis: Check whether the dimensions are correct.

\part b) Consider limiting cases (for example, $\theta=0$).  Does the
range formula work in these cases?

\part c) Give a physical argument for why the range contains a factor
of $v^2$ (instead of, say, simply $v$ or $1/v$ or no $v$ at all).
(Dimensional analysis, which you did in part a, is a mathematical
argument; in this part, you are asked to reinforce the mathematics
with a physical argument.)

\part d) Give a physical argument for the factor of 2.

\part e) Give a physical argument for the $1/g$ factor.

\part f) To derive \eqref{range}, you have to neglect many effects
(for example air resistance).  List as many of these effects as you
can.  Let your imagination run; no effect is too small to mention
here.

\nq Number sense

Without a calculator, estimate

\part a) $\sqrt{1.3}$

\part b) $\root 3 \of {1.6}$

\part c) $\sin 7$

\part d) $1.01^{100}$ (Hint: What is $\ln 1.01$?)

\nqlabel{prob:scales}Scales

You stand on a scale holding a book.  You then place the book next to
you on the scale.  The two scale readings are of course identical.  Of
course!?  Prove it by using Newton's laws and drawing free-body
diagrams.  Clearly label the third-law pairs (pairs that must be equal
and opposite as a consequence of Newton's third law),\ft{I
avoid using the perhaps more familiar term `action--reaction pairs'
because it needlessly confuses; it implies, wrongly, that one force
causes the other.} and describe each force in words.

\nqlabel{prob:pendulum}Pendulum

The figure shows a pendulum at five points in its swing; positions A
and E are the extremes of the motion.  On each bob, draw
(approximately) the acceleration vector at that point in the swing.
If the acceleration is zero (in which case there is no arrow to draw),
simply circle the bob.

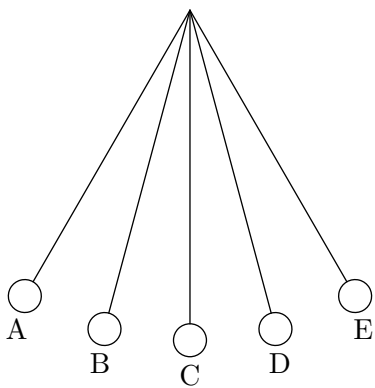
\myfig{fig.6}

\nq Tetrahedron

In methane, the four hydrogen atoms lie at the corners of a regular
tetrahedron, and the carbon atoms lies at the centre.  Where is the
centre of a tetrahedron with unit edge length?  What is the bond angle
(the angle between two C--H bonds)?  (Hint: Make an analogy.)

\nq Estimation

\part a) Estimate how thick a layer of rubber is deposited on the road
by a car tyre.  Comment on your result.

\part b) Estimate the world-record speed for cycling
and for swimming.  (Hint: First estimate how much mechanical power an
athlete can put out.)

\nqlabel{prob:scales-again}Scales (again)

You stand on a scale holding a book (for simplicity of diagramming,
you balance it on your head).  You then place the book next to you on
the scale.  The two scale readings are of course identical.  Of
course!?  Prove the equality by using Newton's laws and drawing free-body
diagrams.  The givens here are your weight and the book's weight.  You
are in effect asked to prove that the weight of the combined you--book
object is the sum of the individual weights.

Draw {\it well-separated\/} free-body diagrams.  Clearly label the
third-law pairs (pairs that must be equal and opposite as a
consequence of Newton's third law); carefully distinguish uses of
Newton's second law from Newton's third law; and describe each force
in words.  Ensure that your argument convinces a skeptical reader
(perhaps try it on your supervision partner), one who says at every
opportunity `Why are those forces equal in magnitude?', `Are you sure
it isn't Newton's third law that justifies this step?', `Or maybe it
should be Newton's second law here?', and so on.

\nqlabel{prob:skiing}Skiing

You (tall rectangle, with mass $m$) stand on a wedge sliding down a
frictionless plane, as shown in the figure.  What weight does the
scale (shaded rectangle) read?  Use clearly labeled, well-separated
free-body diagrams.

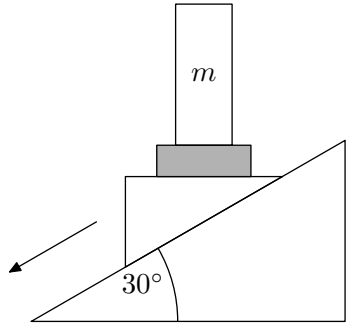
\myfig{fig.7}

\nq Analogy

Into how many regions can $n$ planes divide space?  Find the maximum
number (what conditions on the planes ensure that the number is a
maximum?).  For example, one plane divides space into two regions; two
planes divide space into at most four regions (but only three if you
are unlucky, and only two if you are really unlucky).  Hint: Play with
the one- and two-dimensional versions of this problem, and then try to
generalize the patterns that you find.

\nq Virtual work

The mass $m_1$ slides down the plane with constant velocity, and $m_2$
rises with constant velocity (see the figure).  Use the principle of
virtual work to find the mass ratio $m_1/m_2$.  We live as usual in
the make-believe world of physics: The plane is frictionless, the
string is massless, and the pulley is massless and frictionless.

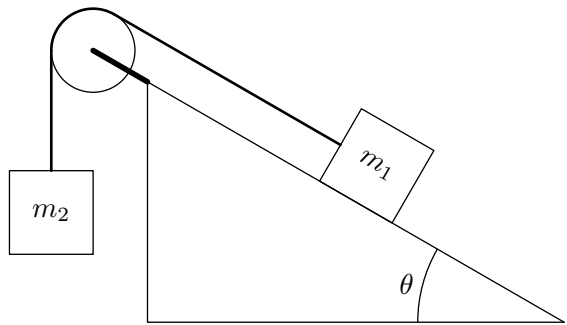
\myfig{fig.8}

\nqlabel{prob:steel-ball} Bouncing ball

You drop a steel ball from a height of one or two metres.  It lands on
a scale and bounces up to nearly the original height.  (Neglect air
resistance.)  Draw free-body diagrams for the ball at four times: (1)
whilst you are holding it, (2) whilst it is falling, (3) when it is
motionless on the scale (namely, just as it starts its upwards
journey), and (4) whilst it is rising.  Indicate qualitatively the
relative magnitudes of the forces.  Sketch qualitatively the scale
reading as a function of time, whilst the ball is on the scale.

\def\mps{\u m \u s^{-1}}
\def\DE{\Delta E}

\nq Improved petrol

Drivers want a petrol that yields more joules per kilogram than
current petrol does.  Discuss the following proposal.

When you accelerate your car from 0 to, say, $15\mps$, the increased
kinetic energy, $\DE_1$, is supplied by burning a quantity of petrol.
Jack is driving in the opposite direction, at $5\mps$.  From his point
of view, your car was going $5\mps$ and, after accelerating, is going
$20\mps$.  He measures a change in kinetic energy, $\DE_2$; and $\DE_2
> \DE_1$ (check this assertion).  The mass of petrol burned is the
same in every reference frame, so Jack measures your petrol to have
more energy per unit mass than you measure it to have.  So, the
proposal is: To increase the energy content of petrol, use a moving
reference frame.

\nqlabel{prob:moments}Moments of inertia

Without evaluating any integrals, rank the following objects in order
of decreasing moment of inertia: (1) a solid sphere, (2) a thin ring,
(3) a spherical shell, and (4) a thin disc.  All objects have the same
mass and radius and are uniform.  For each object, the axis of
rotation passes through the centre of mass.  For the disc and the
ring, the axis is perpendicular to the plane that contains the disc or
ring.  Explain your rankings.

\nqlabel{prob:skiing-pendulum}More skiing

You ski down hill A and up hill B, then ski backwards down hill B and
backwards up hill A (see figure).  There is no friction or air
resistance, so the cycle repeats forever and ever.  Being a skilled
skier, you don't need to clutter your hands with poles; instead, from
your hand, you dangle a string with a mass at its end.  Draw the
direction of the string: (1) as you ski down hill A (square with 1 in
it), (2) as you ski up hill B (square with 2 in it), (3) when you are
momentarily stopped on hill B (square with 3 in it), and (4) as you
ski backwards down hill B (square with 4 in it).  There is plenty of
friction in the oscillations.  How does each string's direction change
if there is slight friction on the slopes?

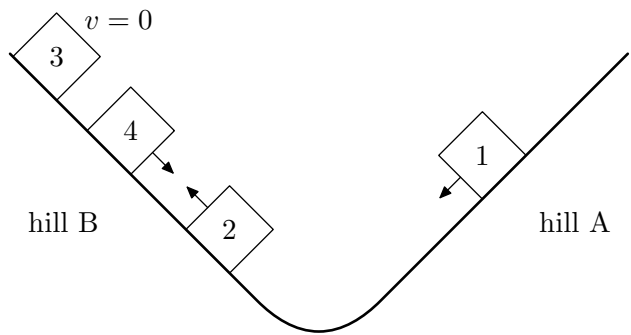
\myfig{fig.9}

\nq Falling moon

The moon is a rock; perhaps large, but it is still a rock.  Why
doesn't it fall to the earth, as other rocks do?  Explain
quantitatively, perhaps with one or two diagrams.

\nq Messy collision

\part i) A ball comes in from the left and causes a series of collisions; the
initial motion is $$
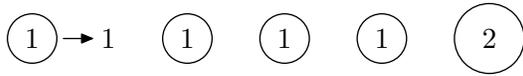
\vbox{\epsfbox{fig.10}}$$
The number in the circle is the object's mass (in arbitrary units) and
the arrow shows the object's velocity (in arbitrary units).  All
motion is one dimensional, and all collisions are elastic.

\def\cfig#1#2{\leftline{\hskip\parindent
${\it #1)}\quad \vcenter{\epsfbox{#2}}$}\par}

Which choice describes the motion after the all the collisions?
\smallskip

\cfig a {fig.11}
\cfig b {fig.12}
\cfig c {fig.13}
\cfig d {fig.14}
\smallskip

\part ii) By transforming to the zero-momentum frame, work out the result of
this collision (also one-dimensional and elastic):
$$
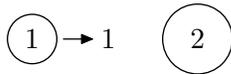
\vbox{\epsfbox{fig.15}}$$
Comment on similarities or differences with part i.

\nq Mathematical conservation

You write a 0 on each vertex of a cube, except for a 1 on one of the
vertices.  Now you play a game.  At each move, you may add 1 to each
of two adjacent numbers (adjacent means connected by an edge).  Your
goal is, using a suitable series of moves, to make all vertex labels
be multiples of 3.  Is this goal possible?  If it is, give the
sequence of moves.  If it is not, prove the impossibility.

\nq Pendulum

As a pendulum slowly loses energy, the amplitude of its swing
decreases.  How does the period change as the amplitude decreases?  Is
it constant, decreasing, or increasing?  Justify your answer.

\nq Centre of mass

A uniform sphere, of radius $r$, has a sphere of radius $r/2$ cut out
of it.  The figure shows a cross section through the sphere.  Where is
its centre of mass?

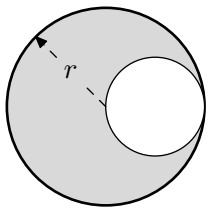
\myfig{fig.16}

\nq Estimation: Oblateness of the earth

Compared to a sphere, the earth is squashed.  

\part a) Why?  Should the polar radius or the equatorial radius be the
larger?

\part b) Which physical
quantities determine $d$, the difference in radii?  How can you
combine these quantities into a length (in other words, into an
estimate for $d$)?

\part c) Use your formula to make a rough numerical estimate of $d$,
and compare it with actual data.

\nq Moments of inertia

\part a) What are the dimensions of moment of inertia?

\part b) An object has mass $M$ and characteristic length
$l$.  The characteristic length is a typical length in the object,
such as a radius or diamater.  What is its moment of inertia, up to a
dimensionless constant?  Consider a geometrically similar object that
is twice as big as this object, in all its dimensions, and made out of
the same material.  What is the ratio of moments of inertia: $I_{\rm
bigger}/I_{\rm smaller}$?

\part c) The moment of inertia of a uniform thin disc is $M\!R^2/2$,
about an axis perpendicular to the plane of the disc and through its
centre.  Perhaps using your results from last week, guess a moment of
inertia for a uniform spherical shell with mass $M$ and radius $R$
(axis of rotation through the centre).  Now calculate it and compare
with your guess.

\nqlabel{prob:rolling}Rolling

Four objects, made of identical steel, roll down an inclined plane.
The objects are (1) a large spherical shell, (2) a large disc, (3) a
small solid sphere, and (4) a small ring.  The large objects have
triple the radius of the small objects.  Rank the objects in order of
decreasing acceleration down the plane.

\nq Buoyancy

A solid iron sphere is floating in a bath of mercury.  You pour water
over the sphere and cover it with water.  Does the sphere rise, sink,
or stay at the same height?

\nq Quadratics by approximation

\part a) Use the quadratic formula and your calculator to find the solutions of
$1 + 200000x+x^2=0$.  What goes wrong?  Why?

\part b) Instead, let's approximate.  If $x$ is near zero, which term can
you neglect?  Solve the simplified equation to get a first
approximation to the smaller root.  Call this first approximation
$x_1$.

\part c) How can you improve your approximation?

\part d) If you know one root, how can you easily find the other root?
\def\w{\omega}
\def\half{{1\over2}}

\nqlabel{prob:snooker}Slipping and sliding

You give a snooker ball (mass $m$ and radius $r$) a horizontal impulse
through its centre of mass and it starts to move with velocity $v_0$.
Let $\mu$ be the coefficient of sliding friction.

\part a) At first, the ball skids; eventually, at some time $t_0$, it
starts to roll.  Why?  On the same graph, sketch qualitatively the
centre-of-mass velocity $v(t)$ and the scaled angular velocity
$r\w(t)$ (rather than $\w$, because $\w$ and $v$ do not have the same
dimensions), label any interesting features, and explain your
reasoning.  Be sure to specify your sign convention for $\w$.

\part b) Qualitatively, how does $t_0$ depend on $m$, $r$, $\mu$,
$v_0$, and $g$?  How should the mass distribution within the ball
affect $t_0$?  (For example, how should $t_0$ for a spherical-shell
ball compare with $t_0$ for a solid-sphere ball?)  Based on your
qualitative reasoning, {\it guess\/} an expression for $t_0$.  Make
sure that your guess has dimensions of time!

\part c) Now analyse the motion quantitatively.  Solve for $v(t)$ and
$r\w(t)$, and sketch them on the same graph.  What is $t_0$?  Compare
with your guess in part b, and discuss any differences.

\part d) What is the final velocity of the ball, $v(t_0)$?  
What is the ball's kinetic energy?  What fraction of its initial
kinetic energy has it lost?

\part e) How much work is done by the force of sliding friction?  Is
your result consistent with the energy loss from part d?

\part f) Try it out: Strike a snooker ball as described, and collect
whatever data you need to make a rough estimate of $\mu$.

\nqlabel{prob:Stirling}Stirling's formula

Stirling's formula says that, for large $n$, 
$$n!\approx \sqrt{2\pi n}\left(n\over e\right)^n.\xeqno{Stirling}$$
Here are two ways to derive a rough version of this formula.

\part a) The first version derives an expression for $\log n!$, which
is also $\sum_{k=1}^n\log k$.  Sketch a graph of $\log k$ and mark the
area represented by the sum $\sum_{k=1}^n\log k$.  As an
approximation, replace the sum by an integral of $\log k$ and evaluate
it to get an approximation to $\log n!$.  Does the integral over- or
underestimate the sum?

\def\zint{\int_0^\infty}

\part b) i) For the second method, begin with a useful trick:
differentiating under the integral sign.  You know that $\zint
e^{-t}\,dt=1$ and, by changing variables, that $$\zint
e^{-at}\,dt={1\over a}.\xeqno{n!-start}$$
Now differentiate both sides of 
this expression $n$ times with respect to $a$, and show that 
$$\zint t^n e^{-t}\,dt = n!.\xeqno{n!-defn}$$

ii) By approximating the integral \eqref{n!-defn}, you can approximate
$n!$.  The integrand is also $e^{f(t)}$ where $f(t)=n\log t-t$.
Sketch $f(t)$ as a function of $t$.  Where is its maximum (call it
$t_0$)?  For large $n$, the exponential of $f(t)$ is even more sharply
peaked than $f(t)$ itself; most of the contribution to the integral
comes from around $t_0$.  Therefore, $n!\sim e^{f(t_0)}$.  What is the
resulting approximation?  How does it compare with Stirling's formula
\eqref{Stirling}?

iii) This last approximation, $n!\sim e^{f(t_0)}$, is dodgy: It
neglects the width of the sharply peaked function $e^{f(t)}$.  A more
accurate approximation is:
$$n!\sim e^{f(t_0)}\x\hbox{width of peak}.$$ Why?  Draw a picture to
explain the argument.  Estimate the width (there are many reasonable
ways to make this estimate) and refine your estimate from ii.  How
does it compare with Stirling's formula?  How could you improve the
approximation yet further?  If you feel adventurous, derive the
$\sqrt{2\pi}$ factor.

\def\dd#1{\langle d_{#1}^2\rangle}

\nq Random walks

A confusing feature of a random walk is the presence of square roots:
Why in a random walk does it take on the order of $N^2$ steps to move
a distance $N$?  Here is one way to understand this bizarre behaviour.
Imagine a particle making a one-dimensional random walk: with equal
probabilities, it moves one step either to the left or to the right.
Let $d_n$ be its position after $n$ steps, with $d_0=0$.  We shall
study $\dd n$, the expected value of $d_n^2$.

\part a) After 0 steps, the distribution of possible $d_0$ is simple:
There is only one possibility, that the particle is at the origin.  So
$\dd0=0$.  After 1 step, the particle is at either $-1$ or $+1$, with
equal probabilities.  So
$$\dd1=\half \left\{(+1)^2 + (-1)^2\right\} = 1.$$ Work out the
probability distribution for the particle position after 2 steps, and
from the distribution, work out $\dd2$.  Repeat for $\dd3$ and $\dd4$.
Generalise the pattern: After $N$ steps, what is the expected squared
distance $\dd N$?  Harder: Prove your result.

\part b) What is $\sqrt{\dd N}$?  Therefore explain the
behaviour mentioned in the introduction.

\nq Air molecules

\part a) Estimate the mean free path, $l$, of air molecules at room
temperature.  This length is the step size in a random walk.

\part b) Roughly how fast does an air molecule move?  Call the speed $c$.

\part c) What dimensions does a diffusion coefficient have?  How can
you combine $c$ and $l$ into a diffusion coefficient?  Estimate the
diffusion coefficient, $D$, for air molecules in air (this coefficient
is called the {\it self-diffusion\/} coefficient of air).  Estimate
how long it would take an air molecule to diffuse across a room.

\part d) Fast pieces of fluid donate momentum to neighbouring slow
pieces of fluid; so the fast pieces slow down, and the slow pieces
speed up.  The viscosity measures the ease with which the momentum
diffuses.  In air, momentum is diffused by particle motion directly:
The particles carry their momentum with them, so viscosity arises from
the same physics as does molecular diffusion.  The viscosity of air
should therefore be related to the diffusion coefficient $D$, which
you estimated in part c.  What are the dimensions of viscosity?  How
can you turn $D$ into a viscosity?  Therefore estimate the viscosity
of air, and compare with reality.  Why can't you use the same method
to estimate the viscosity of water?

\nq Atmosphere thickness

Here is a crude method to estimate the height, $H$, of the earth's
atmosphere.  The atmosphere does not end abruptly at $H$; rather, the
density falls gradually to zero.  You can think of $H$ as the height
at which the density has fallen by a significant fraction.  To
determine $H$, mentally launch an air molecule vertically upwards; how
high does it reach (if there is no atmosphere in its way)?  The height
of course depends on the launch velocity.  How can you choose a
reasonable launch velocity?  Get a numerical estimate for the height.

\nqlabel{prob:isothermal-atmosphere}Atmosphere, take 2

You can also use a more honest method to work out the density versus
height in the atmosphere.  Assume that the atmosphere has a uniform
temperature.  Now work out how the density varies with height.  (Hint:
Consider also how the pressure must vary, and use the ideal gas law to
relate pressure and density.)  Your density should have the form of
the Boltzmann distribution.  Coincidence?  Discuss.

\nq Return probability in random walks

From last week: In a one-dimensional random walk, the particle's rms
distance from the origin after $n$ steps is $\sqrt n$.  You can use
this result to determine the probability that the particle returns to
the origin (the other possibility is that the particle escapes to
infinity and never returns).  The particle's position is distributed
with approximately a Gaussian distribution; the standard deviation is
the rms distance $\sqrt n$.  

Approximate the distribution instead as a rectangle of width $\sqrt
n$.  In other words, replace the Gaussian distribution by a uniform
distribution.  So $p_n$, the probability that the particle is at the
origin after $n$ steps, is $1/\sqrt{n}$ (give or take a constant).
What is the expected number of visits to the origin over all time?
What therefore is the probability that the particle returns to the
origin?  

Extend the argument to two- and three-dimensional random walks.  What
if anything changes as you go from one to two to three dimensions?

\def\1{{\tt 1}}

\nq Tricky die (from vac problems)

You roll a 1000-sided die once per second.

\part a) How long, on average, between rolls of a \1?  \ans $1000\s$

\part b) Your friend Jane walks up and sees you rolling the die.  How
long does she have to wait, on average, before a \1\ turns up?  \ans
$1000\s$ (careful of the gambler's fallacy)

\part c) How long, on average, between the time that she walked up to
you and the time that you {\it last\/} rolled a \1?  \ans $1000\s$

Combining the answers to parts b and c, we conclude that a \1\ turns
up every $2000\s$, in contradiction to part a.  How can you resolve
the paradox?

In the kinetic theory, you find the same paradox.  A molecule
travels on average a distance $l$ (the mean free path) before
colliding with another molecule.  Observe one of the molecules and be
puzzled.  How far away, on average, is its next collision?  \ans
$l$, because molecules have no memory.  How far away, on average, was
its last collision?  \ans $l$, because molecules have no memory.
So the mean free path should be $2l$.

\def\d{\delta}
\def\tc{t_{\rm c}}
\def\td{t_{\rm d}}
\def\vs{c_{\rm s}}
\def\lbar{{\mathchar'26\mkern-9mu\lambda}}
\def\k{\kappa}
\def\Hz{\u{Hz}}

\nq Singing logarithms

Read \xref{page:logarithms}
approximating logarithms, and use the
method to compute $3^8$ and $\log_{10} 5$.  How accurate are the
values?  Make up four more computations in which logarithms would aid
the computation; use the method to do the computations.

\nq Adiabatic or isothermal sound waves?

Newton was the first to work out the speed of sound.  He found that
$\vs =
\sqrt{P/\rho}$.  Today we would deduce the speed by deriving and
solving the wave equation, which is a partial differential equation
for the pressure $p(x,t)$.  When Newton derived the speed, regular
derivatives were barely understood and partial derivative were
unimagined.

Newton's formula implicitly assumes the compressions and rarefactions
that constitute a sound wave are isothermal.  (An adiabatic
compression happens too quickly for heat to flow and thereby to
equalise the temperature with the neighbouring rarefaction.)  Are the
compressions or rarefactions isothermal or adiabatic?

\part a) To decide, consider a sound wave with angular frequency $\w$, which is
$f/2\pi$.  (Angular frequency usually makes for more accurate
estimates than regular frequency does.)  Roughly how long does a
compression last?  Call this time $\tc$.  The size of the compression
region is roughly $\vs/\w$, which is usually called $\lbar$.  Roughly
how long does it take the heat in this region to diffuse outside this
region?  Call this time $\td$.  Hint: In a gas, the
molecular-diffusion constant $D$ is roughly equal to the
heat-diffusion constant $\k$.  Sketch $\tc$ and $\td$ as functions of
$\w$ (on the same graph).  What is special about the intersection
frequency (call it $\w_0$)?  What is $\w_0$ as a function of $\k$ and
$\vs$?

\part b) From Set 6, Question 4d, you know that $\k\sim l\vs$, where $l$ is the
mean free path.  Actually, you found that $D\sim l c$, where $c$ is a
typical molecular speed, but $c\sim\vs$, and $D\sim \k$.  If $\tau$ is
the mean free time, then $l\sim\vs\tau$.  Use this relation to
simplify your expression for $\w_0$.

\part c) Is a sound wave of frequency $f=256\Hz$ adiabatic (this tone
is roughly middle C)?  Therefore decide whether Newton's implicit
assumption is correct.

\part d) Now decide experimentally. Compute $\vs$ for air at sea level
using Newton's formula.  The adiabatic speed is given by
$$\vs^{\rm adiabatic} = \sqrt{\gamma}\, \vs^{\rm isothermal},$$ where
$\gamma$ is the ratio of specific heats $c_{\rm p}/c_{\rm v}$, which
is roughly $1.4$ for dry air.  How closely do the two speeds match the
actual speed of sound?

\nqlabel{prob:teacup-spindown}Teacup spindown

You stir your afternoon tea to mix the milk (and sugar if you have a
sweet tooth).  Once you remove the stirring spoon, the rotation starts
to slow.  What is the spindown time $\tau$?  In other words, how long
before the angular velocity of the tea has fallen by a significant
fraction?

To estimate $\tau$, consider a physicist's idea of a teacup: a
cylinder with height $L$ and diameter $L$, filled with liquid.  Why
does the rotation slow?  Tea near the edge of the teacup -- and near
the base, but for simplicity neglect the effect of the base -- is
slowed by the presence of the edge (the noslip boundary condition);
the edge produces a velocity gradient.  Because of the tea's
viscosity, the velocity gradient produces a force on any piece of the
edge; this force tries to spin the piece in the direction of the tea's
motion.  The piece exerts a force on the tea, which is equal in
magnitude and opposite in sense: The edge slows the rotation.

\part a) In terms of the total viscous force $F$ and of the initial
angular velocity $\w$, estimate the spindown time.  Hint: Consider torque and angular
momentum.  (Feel free to drop any constants, such as $\pi$ and $2$, by
invoking the Estimation Theorem: $1=2$.)

\part b) You can estimate $F$ with the idea that 
$$\hbox{viscous force}\sim \rho \nu \x \hbox{velocity gradient} \x
\hbox{surface area}.\xeqno{viscous-force}$$ Here $\rho \nu$ is $\eta$.
The more familiar viscosity is $\eta$, known as the dynamic viscosity.
The more convenient viscosity is $\nu$, the kinematic viscosity.  (To
see why $\nu$ might be more convenient than $\eta$, work out the
dimensions of $\nu$.)  The velocity gradient is determined by the size
of the region in which the the edge has a significant effect on the
flow; this region is called the boundary layer.  Let $\d$ be its
thickness.  Estimate the velocity gradient near the edge, and use
\eqref{viscous-force} to estimate $F$.

\part c) Put your expression for $F$ into your earlier estimate for
$\tau$, which should now contain only one quantity that you have not
yet estimated (the boundary-layer thickness).

\part d) You can estimate $\d$ using your knowledge of
random walks.  The boundary layer is a result of momentum diffusion;
just as $D$ is the molecular-diffusion coefficient, $\nu$ is the
momentum-diffusion coefficient.  In a time $t$, how far can momentum
diffuse?  This distance is $\d$.  What is a natural estimate for $t$?
(Hint: After rotating 1 radian, the fluid is moving in a significantly
different direction than before, so the momentum fluxes no longer
add.)  Therefore estimate $\d$.

\part e) Now put it all together: What is $\tau$?  

\part f) Stir some tea and estimate $\tau_{\rm exp}$.  Compare this value with
the value predicted by your theory.  In water (and tea is roughly
water), $\nu\sim\e{-6}\m^2\sec^{-1}$.

\nq Stokes' law

You can use ideas from the previous problem to derive Stokes' formula
for drag at low speeds (more precisely, at low Reynolds' number).
Many weeks ago, we derived the result from dimensional analysis; here
you will find a physical argument.  

Consider a sphere of radius $R$ moving with velocity $v$.
Equivalently, in the reference frame of the sphere, the sphere is
fixed and the fluid moves past it with velocity $v$.  Next to the
sphere, the fluid is stationary.  Over a region of thickness $\d$ (the
boundary layer), the fluid velocity rises from zero to the full flow
speed $v$.  Assume that $\d\sim R$ (the most natural assumption) and
estimate the viscous drag force.  Compare the force with Stokes'
formula (remember that $\rho\nu=\eta$).

\vfill\eject
\xrdef{page:logarithms}
\begingroup
\ninepoint

\let\mus=\tenmus
\def\sharp{\rlap{\raise0.4ex\hbox{\mus\char"5D}}}
\def\flat{\rlap{\raise0.2ex\hbox{\mus\char"5B}}}

\def\M{\hbox{M}}
\def\P{\hbox{P}}
\def\m{\hbox{m}}
\def\d{\hbox{d}}
\def\dot{\cdot}
\def\tablerule{\noalign{\vskip2pt}
\noalign{\hrule}
\noalign{\vskip2pt}}

\def\boxit#1{\vtop{\hrule\hbox{\vrule\kern3pt
  \vbox{\kern3pt#1\kern3pt}\kern3pt\vrule}\hrule}}

\newbox\keybox
\newbox\databox

\setbox\keybox=\boxit{\vbox{\halign{
\hfil # \hfil \quad & \hfil # \hfil \quad & \hfil # \hfil \cr
\multispan3 \hfil \bf KEY \hfil\cr
\tablerule
{\sl Symbol} & {\sl Interval} & {\sl Notes}\cr
\noalign{\vskip2pt}
M2 & Major 2nd & C--D\cr
m3 & Minor 3rd & C--E\flat\cr
M3 & Major 3rd & C--E\cr
P4 & Perfect 4th & C--F\cr
d5 & Diminished 5th & C--G\flat\cr
P5 & Perfect 5th & C--G\cr
m6 & Minor 6th & C--A\flat\cr
M6 & Major 6th & C--A\cr
P8 & Octave & C--C\cr
}}}

\setbox\databox = \boxit{\vbox{\openup2pt\halign{
\hfil \hskip3em# \qquad & \hfil $#$ \hfil \qquad & \hfil  $#$
\qquad & \qquad $#$ \hfil\quad\cr
\omit \hfil \hbox{\sl Semitones}\hfil & \hbox{\sl Interval} &
\hbox{\sl Ratio} &\omit
\hfil \hbox{\sl Exact Value}\hfil\cr
\tablerule
2 & \M2 & 9/8 & 1.122 \cr
3 & \m3 & 6/5 & 1.1885\cr
4 & \M3 & 5/4 & 1.259\cr
5 & \P4 & 4/3 & 1.3335\cr
6 & \d5 & \sqrt2 & 1.4125\cr
7 & \P5 & 3/2 & 1.496\cr
8 & \m6 = \P8 - \M3 & 8/5 & 1.585 \cr
9 & \M6 = \P8 - \m3 & 5/3 & 1.679 \cr
10 & \P5 + \m3 & 9/5 & 1.7783\cr
   & 2\dot\P4 & 16/9 & 1.7783\cr
11 &  &         17/9 & 1.8836\cr
12 & \P8 & 2 & 1.9953\cr
17\rlap{.4} & & e& 2.718\cr
19 & \P8 + \P5 & 3 & 2.9854\cr
24 & 2\dot\P8 & 4 & 3.981 \cr
28 & 2\dot\P8 + \M3 & 5 & 5.012\cr
31 & 2\dot\P8 + \P5 & 6 & 5.9566\cr
34 & 3\dot\P8 - \M2  & {64\over9}\approx7 & 7.080\cr
36 & 3\dot\P8 & 8 & 7.943\cr
38 & 2\dot(\P8+\P5) & 9 & 8.913\cr
40 & 3\dot\P8 + \M3 & 10\rlap{.} & \llap{1}0. \cr
}}}

\noindent{\medbf Approximate Logarithms Using
Half-Decibels}\footnote{\raise0.33ex\hbox{$^*$}}%
{Method due to the statistician I.~J.~Good,
who credits his father.}
\medskip

\line{\copy\databox \hfil \copy\keybox}
\bigskip

\def\two{2^{1/12}}
\def\ten{10^{1/40}}

\noindent
The starting point is $2^{10}\approx10^3$, or $\two\approx\ten$.  By
chance $\two$ is the semitone frequency ratio on the equal-tempered
scale.  Since we know what Pythagorean ratios the equal-tempered
intervals are supposed to approximate, we can approximate logarithms
to the base $\two$, and thereby approximate logarithms to the base
$\ten$.  The ratio column indicates the ratios for perfect Pythagorean
intervals, and the exact value column shows $10^{{\rm semitones}/40}$,
to show the accuracy of the method.  Note that 10 semitones has two
possible breakdowns into intervals, as $\P5+\m3$ or $2\dot\P4$.  The
second is much more accurate, because in the equal-tempered scale, the
perfect intervals come out almost exactly right, at the cost of some
error in the major and minor intervals.

To use the table to compute $\log_{10} x$, find $x$ as a product of
ratios, add the number of semitones for the ratios, and divide by 40
(divide by 2 to get dB).  To calculate $10^x$, multiply $x$ by 40,
find that value in the semitones column, and read off the
corresponding ratio.  From a few basic Pythagorean ratios and number
of semitones, most of the table is easy to figure out.  The most
important to remember one is the fifth: 7~semitones corresponds
to~3/2.  For example, from the fifth we can compute the frequency
ratio for a fourth (5~semitones).  The two intervals together make an
octave, so the product of their frequency ratios is~2.  This
means 5~semitones corresponds to 4/3.  Many other entries can be
worked out similarly.

\def\from{\leftarrow}
\def\semi{\,\hbox{semitones}}
\def\dec{\,\hbox{decades}}
\def\dB{\,\hbox{dB}}
\def\octave{\,\hbox{octave}}
Some examples (arrows point from the real to the log world):
 $$2\to1\octave=12\,\semi=6\dB=0.3\dec.$$
$$\left({4\over3}\right)^{10}\to 10\dot\P4=50\,\semi=40 + 2\dot\P4\from
10\cdot{16\over9} = 17.78 \hbox{ (exact 17.76)}.$$
$$5={5\over4}\dot2\dot2\to\M3+2\dot\P8=28\semi={28\over40}\hbox{ or }0.7\dec\ 
(14\dB).$$
\endgroup

\end